\documentclass[a4paper]{article}

\usepackage{INTERSPEECH2022}

\usepackage{subcaption}
\usepackage{enumitem}
\setlist[itemize]{align=parleft,left=0pt..1em}
\usepackage{hyperref}
\hypersetup{
    colorlinks=true,
    linkcolor=blue,
    filecolor=magenta,      
    urlcolor=cyan,
    citecolor=blue,
}
\usepackage{colortbl}
\usepackage{xcolor}
\definecolor{Dejan}{rgb}{1.0,0.0,0.0}
\definecolor{israel}{rgb}{1.0,0.5,0.0}
\definecolor{wenchin}{rgb}{1.0,0.5,0.5}

\usepackage{tikz}  
\usepackage{cite}

\expandafter\def\expandafter\normalsize\expandafter{%
    \normalsize
    \setlength\abovedisplayskip{5pt}
    \setlength\belowdisplayskip{5pt}
    \setlength\abovedisplayshortskip{5pt}
    \setlength\belowdisplayshortskip{5pt}
}

\def\l2{{\ell_2}}

\title{ End-to-End Binaural Speech Synthesis}
\name{Wen-Chin Huang$^{1*}$\thanks{$^{*}$Work done while interning at Meta Reality Labs Research.}, Dejan Markovi\'c$^2$, Israel D. Gebru$^2$, Anjali Menon$^2$, Alexander Richard$^2$}
\address{
  $^1$Nagoya University, Japan\\
  $^2$Meta Reality Labs Research, USA}
\email{wen.chinhuang@g.sp.m.is.nagoya-u.ac.jp\\
      \{dejanmarkovic,idgebru,aimenon,richardalex\}@fb.com}

\begin{document}

\maketitle
\begin{abstract}
In this work, we present an end-to-end binaural speech synthesis system that combines a low-bitrate audio codec with a powerful binaural decoder that is capable of accurate speech binauralization while faithfully reconstructing environmental factors like ambient noise or reverb. The network is a modified vector-quantized variational autoencoder, trained with several carefully designed objectives, including an adversarial loss. We evaluate the proposed system on an internal binaural dataset with objective metrics and a perceptual study. Results show that the proposed approach matches the ground truth data more closely than previous methods. In particular, we demonstrate the capability of the adversarial loss in capturing environment effects needed to create an authentic auditory scene.
\end{abstract}

\section{Introduction}
\label{sec:intro}

Augmented and virtual reality technologies promise to revolutionize remote communications by achieving spatial and social presence, i.e., the feeling of shared space and authentic face-to-face interaction with others. High-quality, \textit{accurately spatialized} audio is an integral part of such an AR/VR communication platform. In fact, binaural audio guides us to effortlessly focus on a speaker in multi-party conversation scenarios, from formal meetings to causal chats \cite{10.1162/pres.1996.5.3.290}. It also provides surround understanding of space and helps us navigate 3D environments.

\begin{figure}[t]
	\centering
	\begin{tikzpicture}
	 \node [below left] (image) at (0,0) {\includegraphics[width=0.97\columnwidth]{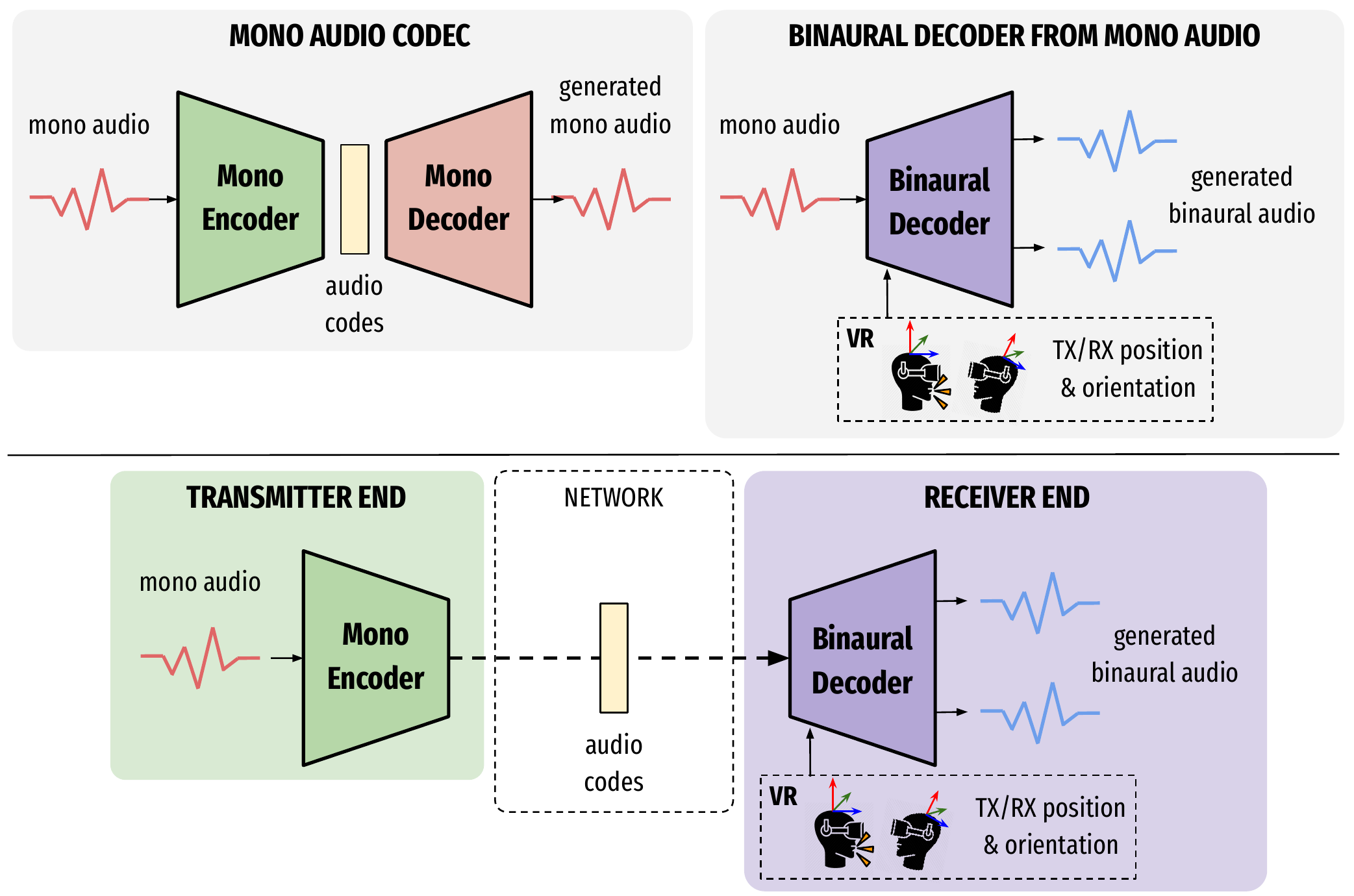}};
	 \node [below left,text width=3cm,align=left] at (-4.7,-4.9){\scriptsize \bf{PROPOSED SYSTEM}};
	 \node [below left,text width=3cm,align=left] at (-4.7,-2.2){\scriptsize \bf{PREVIOUS WORKS}};
    \end{tikzpicture}
	\caption{Illustration of previous works and the proposed system. \textbf{Top left}: Standard audio codec which encodes and reconstructs mono audio. \textbf{Top right}: binaural decoder that spatializes mono audio by conditioning on orientation and relative position between the transmitter and receiver. \textbf{Bottom}: proposed end-to-end binaural system that combines previous modules. \label{fig:framework}}
	\vspace{-0.5cm}
\end{figure}

Our goal is to create a pipeline for a binaural communication system, as shown at the bottom of Fig.~\ref{fig:framework}.
At the transmitter end, monaural audio is first encoded by an audio encoder, and then transmitted over the network.
At the receiver end, the transmitted audio code is decoded, and the binaural audio is synthesized according to transmitter and receiver positions in the virtual space.
Specifically, such a system should be capable of \textit{(a)} encoding transmitter audio into a \textit{low-bitrate neural code} and \textit{(b)} synthesizing binaural audio from these codes \textit{including environmental factors} such as room reverb and noise floor, which are crucial for acoustic realism and depth perception.

Although binaural synthesis has recently experienced a breakthrough based on neural audio rendering techniques~\cite{binaural-decoder,implicit-HRTF,richard2022deepimpulse} that allow to learn binauralization and spatial audio in a data-driven way, these approaches fall short in their ability to faithfully model environmental factors such as room reverb and noise floor. The reason these models fail to model stochastic processes is their reliance on direct reconstruction losses on waveforms. Additionally, this reliance on metric losses makes the joint optimization of neural spatial renderers and neural audio codecs a difficult task.
In fact, given their high sensitivity to phase shifts that do not necessarily correlate with perceptual quality, metric losses are known to perform badly in pure generative tasks, including speech synthesis from compressed representations. 
Yet, efficient compression and encoding are required in a practical setting like an AR/VR communication system.

In this work, we demonstrate that these shortcomings of existing binauralization systems can be overcome with adversarial learning which is more powerful at matching the generator distribution with the real data distribution.
Simultaneously, this paradigm shift in training spatial audio systems naturally allows their fusion with neural audio codecs for efficient transmission over a network.
We present a fully end-to-end, waveform-to-waveform system based on a state-of-the-art neural codec \cite{soundstream} and binaural decoder \cite{binaural-decoder}. The proposed model borrows the codec architecture from \cite{soundstream} and physics-inspired elements, such as view conditioning and time warping, from \cite{binaural-decoder}. We propose loss functions and a training strategy that allows for efficient training, natural sounding outputs and accurate audio spatialization.
In summary, our contributions are as follows:
\begin{itemize}
    \itemsep 0em 
    \item we propose a first fully end-to-end binaural speech transmission system that combines low-bitrate audio codecs with high-quality binaural synthesis;
    \item we show that our end-to-end trained system performs better than a baseline that cascades a monaural audio codec system (top left of Fig.~\ref{fig:framework}) and a binaural decoder (top right of Fig.~\ref{fig:framework});
    \item we demonstrate that adversarial learning allows to faithfully reconstruct realistic audio in an acoustic scene, including stochastic noise and reverberation effects that existing approaches fail to model.
    
    
\end{itemize}

\begin{figure*}[t]
	\centering

	\begin{subfigure}{1.28\columnwidth}
	    \centering
	    \includegraphics[width=\textwidth]{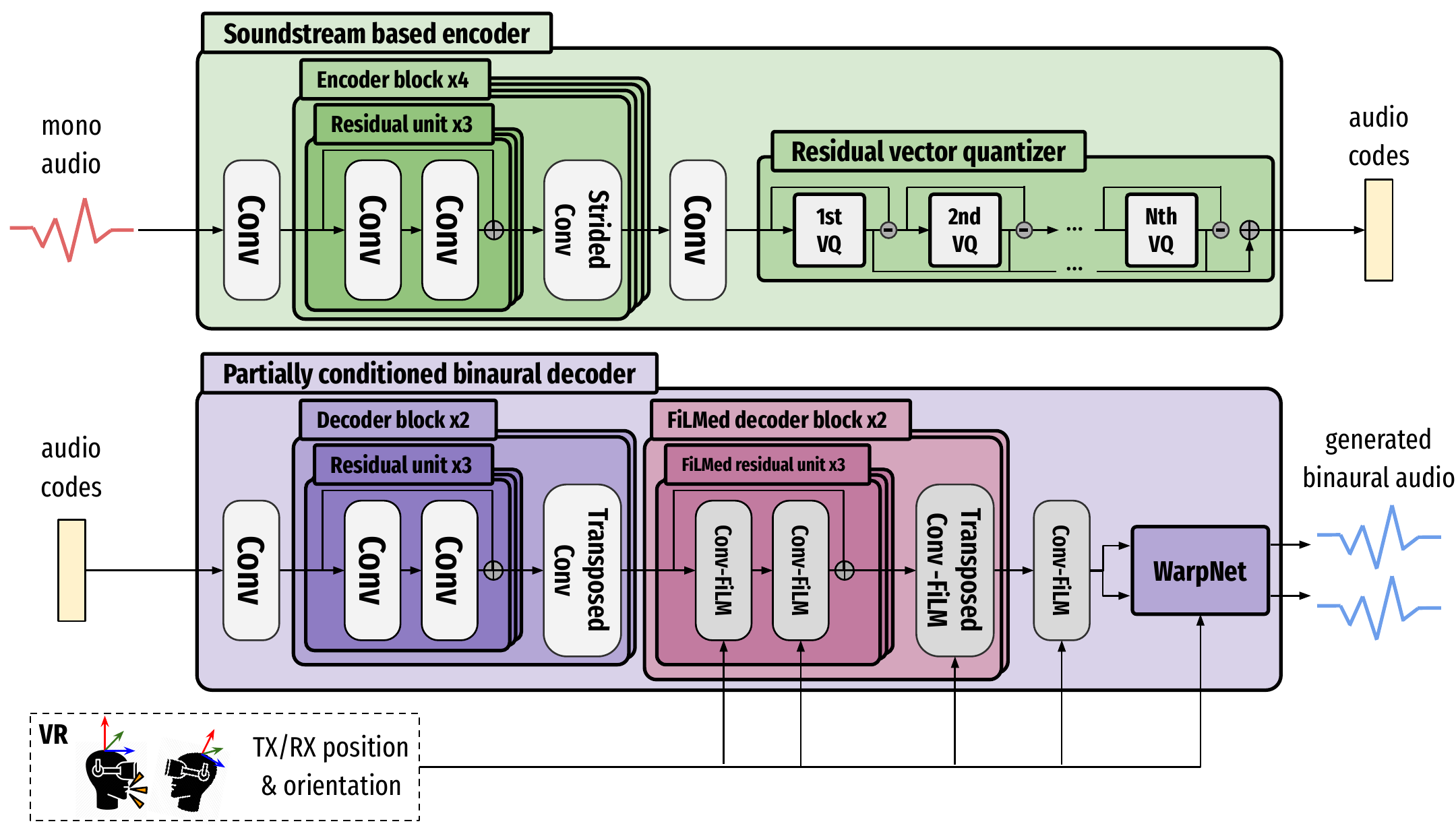}
	\end{subfigure}
	\hspace{5pt}
	\begin{subfigure}{0.72\columnwidth}
	    \centering
	    \includegraphics[width=\textwidth]{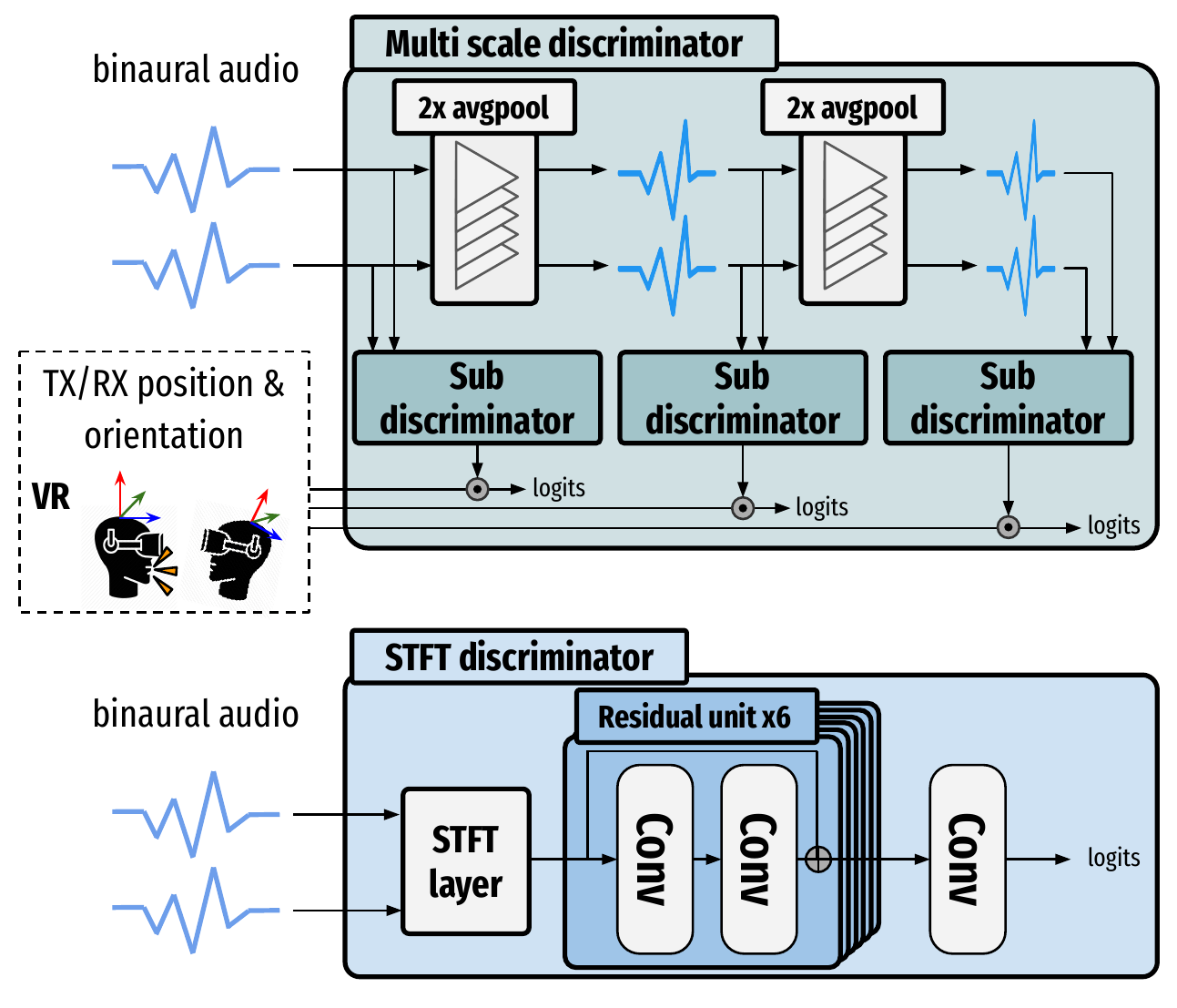}
	\end{subfigure}
	
	\caption{Model architecture. \textbf{Top left}: Soundstream-based encoder, consists of a stack of conv layer-based encoder blocks and a residual vector quantizer. \textbf{Bottom left}: Partially conditioned binaural decoder, consists of a stack of partially FiLMed decoder blocks and a WarpNet. \textbf{Top right}: Multi-scale projection discriminator. \textbf{Bottom right}: STFT discriminator. \label{fig:model}}
	\vspace{-0.3cm}
\end{figure*}

\section{Related Work.}
\textbf{Audio codecs} have long relied on traditional signal processing and in-domain knowledge of psychoacoustics in order to perform encoding of speech \cite{LPC} or general audio signals \cite{opus, EVS}. More recently, following advances in speech synthesis \cite{wavenet, melgan, hifigan}, data-driven neural audio codecs were developed \cite{wavenet-speech-coding, speech-coding-vqvae-wavenet, zhen2020cq, Lyra, polyak21_interspeech}, and Soundstream~\cite{soundstream}, a novel neural audio codec, has shown to be capable of operating on bitrates as low as 3kbps with state-of-the-art sound reconstruction quality.
None of these approaches, however, was developed with spatial audio in mind, focusing solely on reconstructing monaural signals, as illustrated at the top left of Fig.~\ref{fig:framework}.

\textbf{Binaural audio synthesis} has traditionally relied on signal processing techniques that model the physics of human spatial hearing as a linear time-invariant system \cite{savioja1999creating, rendering-localized, natural-sound-rendering,  surround-by-sound}. More recently, there has been a line of studies on neural synthesis of binaural audio that showed the advantages of the data-driven approaches \cite{ssl-spatial-audio, 2-5d, ssl-spatial-classifier, Yang_2020_CVPR, zhou2020sep, binaural-decoder, implicit-HRTF, richard2022deepimpulse}.
We will refer to these models, illustrated at the top right of Fig.~\ref{fig:framework}, as \textit{binaural decoders}.
All these approaches, however, are trained as regression models with point-wise, metric losses such as mean squared error. Consequently, they fail to model stochastic processes on the receiver side that are not observable on the mono transmitter input. Examples of these processes are noise floor and reverberant effects in the virtual receiver environment.

\section{Proposed system}

\subsection{Model architecture}

Formally, we aim to find a model $f$ that takes as input a mono audio signal $x\in\mathbb{R}^T$, and generates the left and right binaural signals $\hat{y}=(\hat{y^{(l)}},\hat{y^{(r)}})$ (both of length $T$) by conditioning on a temporal signal $\bm{c}$ of length $T$ which contains the transmitter and receiver position and orientation.
Our model, depicted in Fig.~\ref{fig:model}, is based on Soundstream \cite{soundstream}, with a series of modifications for generating binaural signals. The input signal $x$ is first encoded with a convolutional (conv) neural network, $\operatorname{Enc}$, and then discretized with a residual vector quantizer (RVQ) to obtain the audio codes, $\bm{h}\in\mathbb{R}^{\frac{T}{M}\times D}$, where $M$ is the downsampling rate and $D$ is the dimension of a single code. The decoder, $\operatorname{Dec}$, which consists of a convnet and a warpnet \cite{binaural-decoder}, then generates the binaural signals by conditioning on the position information. The process can be formulated as follows: 
\begin{equation}
    \hat{y^{(l)}},\hat{y^{(r)}}=f(x, \bm{c})=\operatorname{Dec}(\bm{h}, \bm{c}),\ \ \bm{h}=\operatorname{Enc}(x).
    \label{eq:formulation}
\end{equation}
In order to facilitate adversarial training, a set of discriminators is trained together with the entire network in an  end-to-end fashion. We describe each component in detail below, and because we mostly followed the specifications described in \cite{soundstream}, we omit detailed hyperparameters due to space constraints.

\subsubsection{Soundstream-based encoder}

The first part of the encoder is a stack of four 1D conv blocks, with each block containing three residual units and a downsampling strided conv layer. After the input mono signal is transformed into a series of continuous vectors, they are thereafter discretized through a RVQ \cite{soundstream, vqvae} of $N$ VQ layers, which represents each vector with a sum of codewords from a set of finite codebooks. These final vectors are denoted as the \textit{audio codes}. Note that in \cite{soundstream}, several techniques are used to improve codebook usage and bitrate scalability including k-means based initialization, codeword revival and quantization dropout, which we did not find necessary in our work. During training, the codebooks are updated with exponential moving averages, following \cite{soundstream, vqvae2}.

\subsubsection{Partially conditioned binaural decoder}

The first part of the decoder is a reverse mirror of the encoder, with the downsampling strided conv layers replaced by upsampling transposed conv layers. Because it was orginally proposed for mono audio reconstruction, we carefully designed the decoder to capture the required fidelity of the binaural signals by conditioning on the position information $\bm{c}$. First, a FiLM-based affine layer \cite{film} was added to the output of each conv layer. Specifically, the position information is first processed through a three-layered MLP with ReLU activation, which is then upsampled to be used as the scale and shift parameters to perform feature-wise affine transformation. Second, due to the low dimension and low frequency nature of the position vector, we further adopt a Gaussian Fourier encoding layer \cite{fourier-mapping} at the beginning of the position input to learn the implicit, high frequency correlation between the position vector and the binaural audio. Moreover, we empirically discovered that it is sufficient to condition only the last few decoder blocks with position information to get high-quality binaural signals. This is because interaural differences are typically determined within a short temporal window ($\le 100$ samples), and the position information is only needed to accurately shift and scale the binaural signals by such a small amount. Since the temporal resolution of the audio codes (determined by the encoder downsampling rate) is greater than this difference, introducing the conditioning at the start of the decoder is ineffective.

Additionally, we added a neural time warping layer proposed in ~\cite{binaural-decoder} at the end of the decoder to model the temporal shifts from mono to binaural signals caused by sound propagation delays. The layer is a fully differentiable implementation of the monotonous dynamic time warping algorithm. 



\subsubsection{Multi-scale and single-scale discriminators}

Following \cite{soundstream}, we used two types of discriminators. The first type is a single-scale STFT discriminator, which operates on the STFT spectrogram. The architecture is based on a stack of conv layer-based residual units. The second type, originally proposed in \cite{melgan}, is a multi-scale discriminator (MSD) with three sub-discriminator operating on different temporal scales: 1/2/4$\times$ downsampled version of the input signal. Each sub-discriminator is composed of a sequence of strided and grouped convolutional layers. In addition, we adopted the projection discriminator proposed in \cite{projection-gan} to inform the multi-scale discriminator to make use of the conditional information when approximating the underlying probabilistic model given in Eq.~\ref{eq:formulation}. We empirically found that this significantly improves the quality of spatialization.

\subsection{Loss function}

Let the target binaural signals be $y=(y^{(l)},y^{(r)})$. Given the importance of interaural time and level differences for human auditory perception~\cite{darwin1999auditory}, we optimize the difference between the left and right predicted signal against the target signal,
\begin{align}
    \vspace{-1cm}
    &\mathcal{L}_{\text{diff}}= \mathbb{E} \Big[ \big\Vert \big(\hat{y^{(l)}}-\hat{y^{(r)}}\big)-\big(y^{(l)}-y^{(r)}\big) \big\Vert_{2} \Big].
    \vspace{-10pt}
\end{align}
We additionally use a phase loss $ \mathcal{L}_\text{pha} $ that directly optimizes the phase in angular space, which has been proven crucial for accurate phase modeling in~\cite{binaural-decoder}.

We also adopted a mix of losses used in \cite{soundstream}. The first loss is a hinge adversarial loss, where the respective losses for the generator (the model $f$ in Eq.~\ref{eq:formulation}) and the discriminator $D$\footnote{For simplicity we assume that $D$ outputs the average logits of all sub-discriminators in this section.} are defined as:
\begin{align}
    &\mathcal{L}_{\text{adv}}^{\text{dis}} =\mathbb{E} \Big[ \max\big(0, 1-D(y, \bm{c})\big)+\max\big(0, 1+D(\hat{y}, \bm{c})\big) \Big] ,\\
    &\mathcal{L}_{\text{adv}}^{\text{gen}} =\mathbb{E} \Big[ \max\big(0, 1-D(\hat{y}, \bm{c})\big)\Big].
\end{align}
Second, the feature matching loss \cite{melgan, hifigan} is introduced as an implicit similarity metric defined as the differences of the intermediate features from the discriminator between a ground truth and a generated sample:
\begin{align}
    \mathcal{L}_{\text{fm}} = \mathbb{E} \Big[ \sum_{l=1}^{L} \lVert D^i(y)-D^i(\hat{y})\rVert_2 \Big],
\end{align}
where $L$ denotes the total number of layers in $D$ and $D^i$ denotes the features from the $i$-th layer. Finally, the mel spectrogram loss is applied, as in \cite{hifigan}:
\begin{equation}
    \mathcal{L}_{\text{mel}} = \mathbb{E} \Big[ \lVert \phi(y)-\phi(\hat{y})\rVert_1 \Big],
\end{equation}
where $\phi$ denotes the transform from audio to mel spectrogram.

The overall generator loss is a weighted sum of the different loss components:
\begin{equation}
    \mathcal{L}^{\text{gen}} = \lambda_\text{diff}\mathcal{L}_{\text{diff}} + \lambda_\text{pha}\mathcal{L}_{\text{pha}} + \lambda_\text{adv}\mathcal{L}_{\text{adv}} + \lambda_\text{fm}\mathcal{L}_{\text{fm}} + \lambda_\text{mel}\mathcal{L}_{\text{mel}}.
\end{equation}

Our initial experiments with weights suggested in \cite{soundstream} ($\lambda_\text{fm}=100, \lambda_\text{mel}=1$) yielded poor results. Instead, we discovered that it is critical to give the mel spectrogram loss a higher weight. The final weight combination we used is: $\lambda_\text{diff}=\lambda_\text{adv}=1, \lambda_\text{pha}=0.01, \lambda_\text{fm}=2, \lambda_\text{mel}=45$.

\subsection{Mono pretraining}


From Eq.~\ref{eq:formulation}, the $\operatorname{Dec}$ is responsible for (1) upsampling $\bm{h}$ to match the temporal resolution of $x$, and (2) spatialization using information from $\bm{c}$. If the model is trained from scratch, the $\operatorname{Dec}$ struggles to achieve both tasks concurrently. As a result, we propose a pretraining strategy, which we found to be important for fast convergence and high-quality output. In the pretraining step the model is trained to generate two copies of the monaural input signals. The primary objective is to train the decoder to upsample while ignoring the condition information via a constant zero vector. Following that, the fine-tuning step is performed using the acutal binaural signals and position condition information. Once the model has been initialised to perform well at upsampling, it can be trained to spatialize and is expected to retain the ability to upsample. 





\section{Experiments}

\subsection{Experimental settings}

\noindent{\textbf{Datasets.}}
We re-recorded the VCTK corpus \cite{vctk} using a binaural microphone setup comprised of three 3Dio Omni Pro rigs, which were placed at the center of a non-anechoic room.
Speech signals were played back on a loudspeaker, which was carried by a person walking randomly around the room to cover various areas.
The 3D position and orientation of the loudspeaker as well as the static 3DIO rigs were tracked using Motive Optitrack system.
We recorded 42 hours of binaural audio data, covering a distance of 4.6 m horizontally and 2.4 m vertically.
The audio was sampled at 48kHz and the tracking data was recorded at 240 frames per second.
For mono-pretraining, we used the original monaural version of VCTK.


\noindent{\textbf{Competing systems.}} We first consider the state-of-the-art binaural \textbf{decoder only} system \cite{binaural-decoder}. It is trained on the same binaural speech dataset. We then consider a \textbf{baseline} system, where we directly cascade the Soundstream \cite{soundstream} and the binaural decoder \cite{binaural-decoder} models trained separately on VCTK and the binaural speech datasets, respectively.

\noindent{\textbf{Objective metrics.}}
The $\l2$ distance of the predicted and ground truth audio is calculated in the waveform and mel spectrogram domains.
To assess the spatialization accuracy, we report the deep perceptual spatial-audio localization metric (DPLM) \cite{DPLM}.

\noindent{\textbf{Subjective evaluation protocol.}}
This evaluation is divided into two parts. 
In the first part, participants were presented the result of our system and a competing system (either the decoder only or the baseline system) and were asked to determine which of them is closer to the ground truth.
The second part focuses on spatialization. The reference and synthetic samples are played alternating, switching between the one and the other every few seconds, so the listeners can observe the change in the sound source position when the switch happens.
Participants are asked which of the synthetic samples has source position closer to the reference. 
Participants annotated more than 350 test examples. 

\subsection{Empirical Evaluation}
\begin{table}[t]
	\centering
	\caption{\textbf{Objective evaluation} results on the competing systems and variations of the proposed systems.}
	
	\centering
	\begin{tabular}{ l c c c }
		\toprule
        System & Wave-$\l2$ $\downarrow$ & Mel spec-$\l2$ $\downarrow$ & DPLM $\downarrow$\\
		\midrule
        Decoder only & 0.228 & 1.220 & 0.108\\
		Baseline & 0.750 & 1.173 & 0.105\\
		Proposed system & 0.807 & 0.631 & 0.106 \\

        
		\bottomrule
	\end{tabular}
	\label{tab:obj-eval}
\end{table}


\begin{figure}[t]
	\centering
    \begin{subfigure}{\columnwidth}
	    \centering
	    \includegraphics[width=\textwidth]{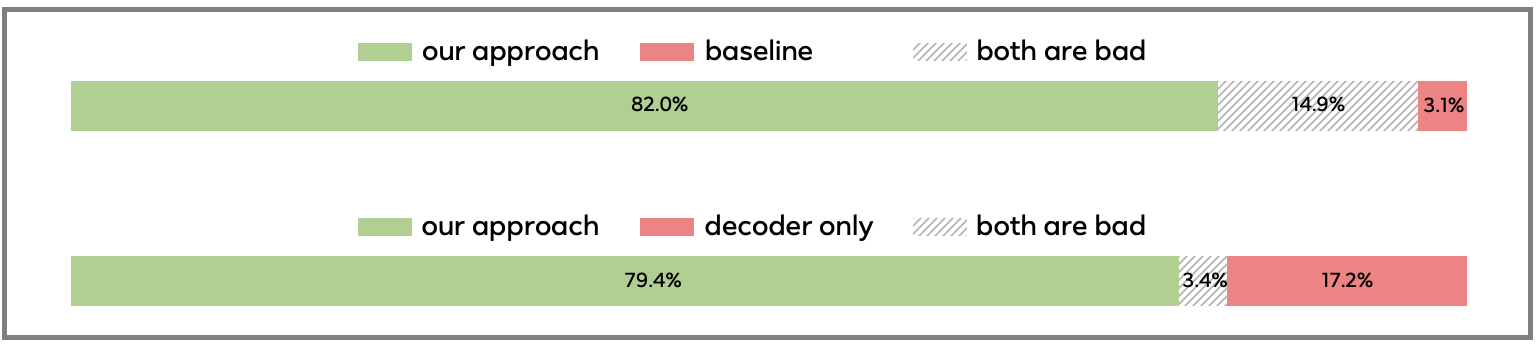}
	    \caption{Evaluation 1: participants were asked if our system or the baseline (top)/decoder only (bottom) system are closer to the ground truth.}
   		\label{fig:sub_q1}
   		\vspace{3mm}
	\end{subfigure}
	
	\begin{subfigure}{\columnwidth}
	    \centering
	    \includegraphics[width=\textwidth]{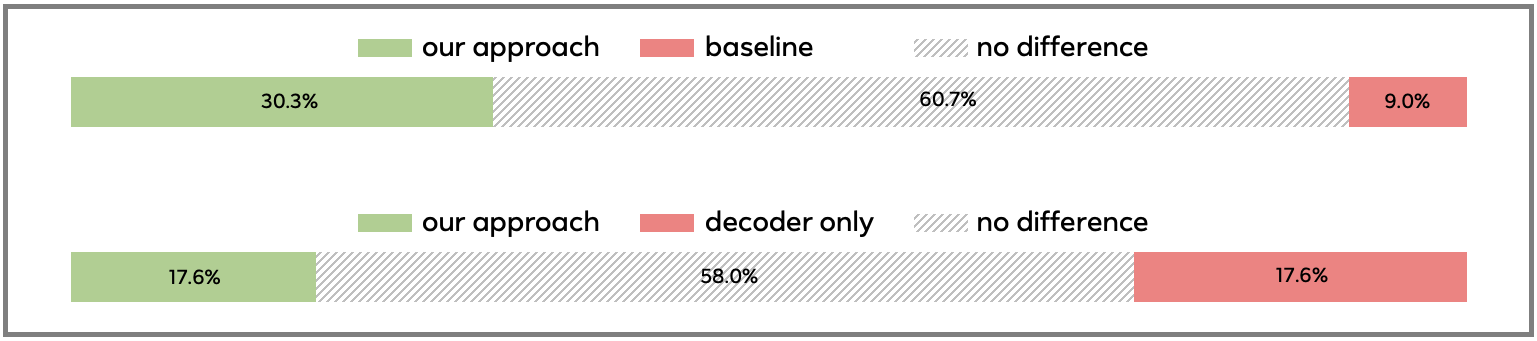}
	    \caption{Evaluation 2: participants were asked if our system or the baseline (top)/decoder only (bottom) system are more accurately spatialized.}
   		\label{fig:sub_q2}
   		\label{fig:sub_q2}
	\end{subfigure}
	\caption{\textbf{Subjective evaluation} results.
	\label{fig:sub}}
	\vspace{-5mm}
\end{figure}

\begin{figure}[t]
	\centering
	\begin{subfigure}{.45\columnwidth}
	    \centering
	    \includegraphics[width=\textwidth]{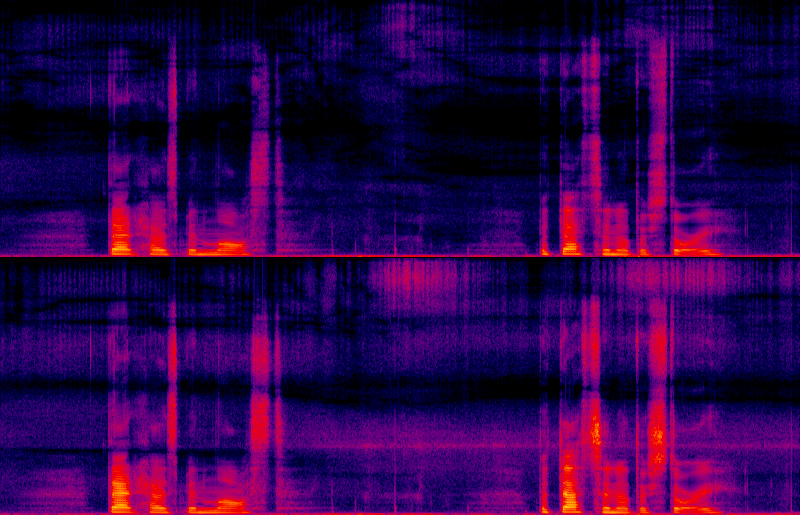}
	    \caption{Decoder only}
   		\label{fig:spec-ub}
	\end{subfigure}%
	\hspace{2pt}
	\begin{subfigure}{.45\columnwidth}
	    \centering
	    \includegraphics[width=\textwidth]{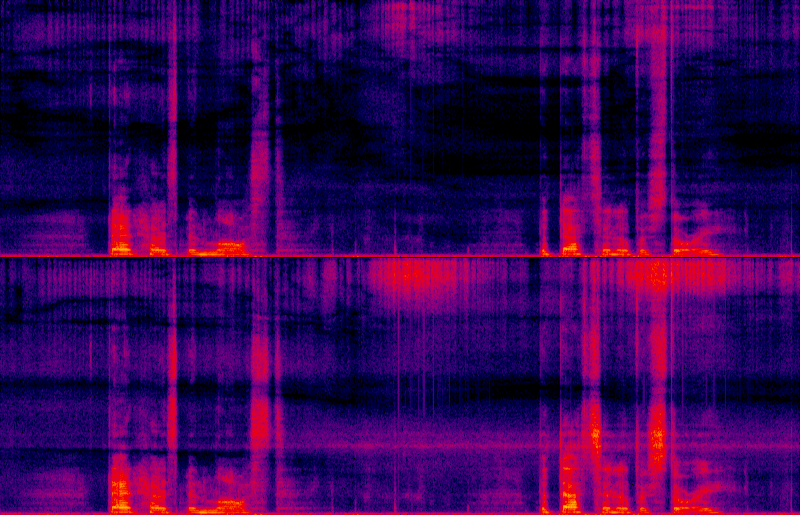}
	    \caption{Baseline}
   		\label{fig:spec-baseline}
	\end{subfigure}
	
	\begin{subfigure}{.45\columnwidth}
	    \centering
	    \includegraphics[width=\textwidth]{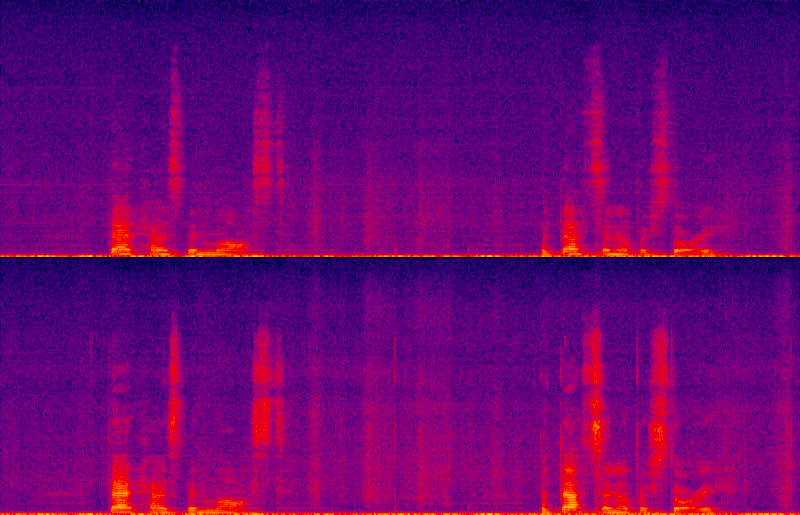}
	    \caption{Proposed system}
   		\label{fig:spec-proposed}
	\end{subfigure}%
	\hspace{2pt}
	\begin{subfigure}{.45\columnwidth}
	    \centering
	    \includegraphics[width=\textwidth]{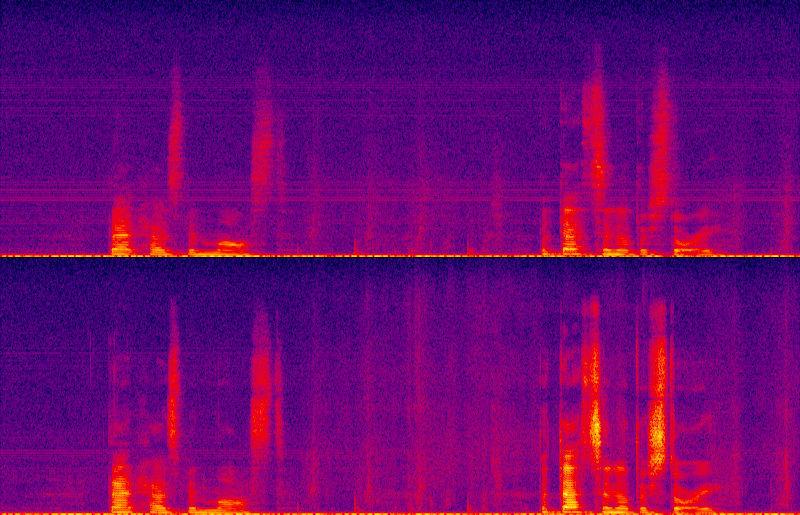}
	    \caption{Ground truth}
   		\label{fig:spec-gnd}
	\end{subfigure}

	\centering
	\caption{Visualizations of spectrograms from the decoder only, baseline, proposed system and the ground truth.}
	\label{fig:spec}
\end{figure}

\textbf{Objective Evaluation.} The objective evaluation results are shown in Tab.~\ref{tab:obj-eval}. 
Unsurprisingly, being optimized in the waveform domain, the decoder only model outperforms others on wave-$\l2$ metric. However, the $\ell_2$-loss is not a good indicator of signal quality and can result in highly distorted signals even if the loss itself is low.
The proposed system is superior in terms of mel spec-$\l2$. We note that the mel spectrogram loss is more indicative of signal quality than the waveform $\ell_2$. In fact, spectrogram visualizations in Fig.~\ref{fig:spec} show that the proposed system matches the ground truth much better than both baseline and decoder only models. 
Finally, the DPLM scores shows that the proposed approach achieves the same spatialization quality as the state-of-the art binaural decoders.


\noindent
\textbf{User study.} The subjective evaluation results are shown in Fig.~\ref{fig:sub}. The first evaluation confirms that the proposed approach generates more natural outputs that are closer to the ground truth recordings than both baseline and decoder-only models. When listening to outputs generated by the baseline and decoder-only models, we found that these models have difficulty reconstructing output that is uncorrelated or only weakly correlated to the input such as room noise floor and reverberation. As a result, these effects are masked out, whereas our approach models them accurately.
The second evaluation confirm that the proposed approach achieves the same level of spatialization quality as the state-of-the art binaural decoders. The results also correlated well with DPLM scores presented in Tab.~\ref{tab:obj-eval}.\footnote{Audio samples can be found at \url{https://unilight.github.io/Publication-Demos/publications/e2e-binaural-synthesis}}

\begin{figure}[h]
	\centering
	\includegraphics[width=0.8\columnwidth]{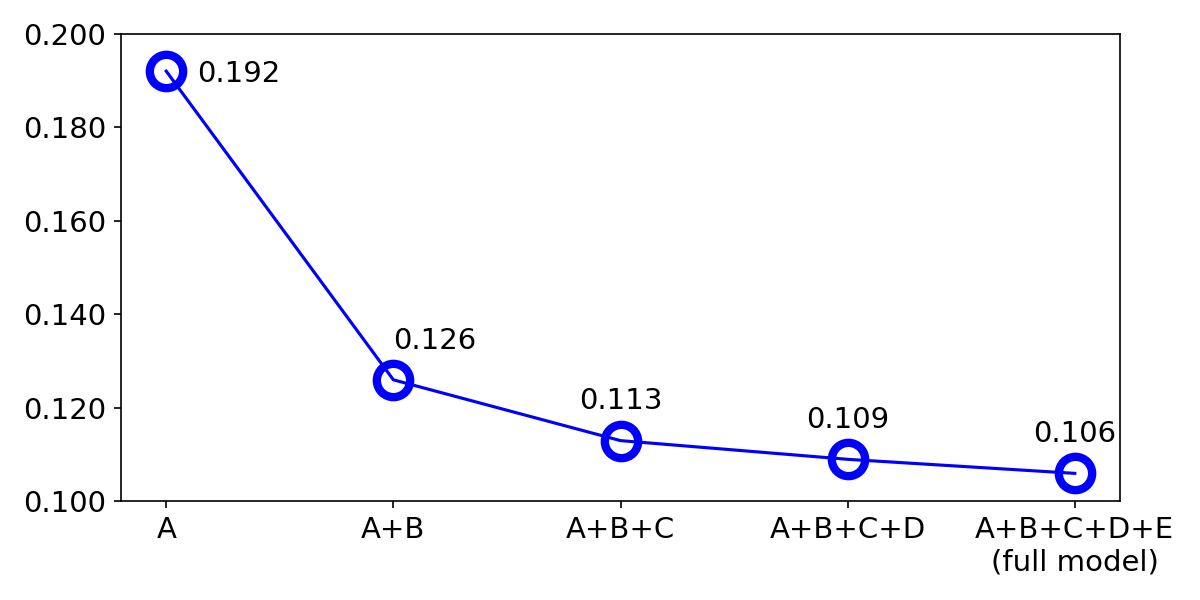}
	\vspace{-3mm}
	\caption{Distances calculated by the deep perceptual spatial-audio localization metric (DPLM) from different variations of the model. Smaller the better. \textbf{A}: mel spectrogram loss ($\mathcal{L}_{\text{mel}}$). \textbf{B}: adversarial-related loss ($\mathcal{L}_{\text{adv}} + \mathcal{L}_{\text{fm}}$). \textbf{C}: mono pretraining. \textbf{D}: partially-conditioned decoder. \textbf{E}: projection discriminator. \label{fig:DPLM-bar}}
	\vspace{-5mm}
\end{figure}

\noindent
\textbf{Ablation study.} We conducted ablation studies to understand the impact of various design choices in the proposed system by gradually adding components and calculating the DPLM distances of different model variations. Results are shown in Fig.~\ref{fig:DPLM-bar}. We see that all model components contribute to the DPLM metric, demonstrating the significance of our design choices.

\noindent
\textbf{Effectiveness of the adversarial loss.}
Note especially the importance of the adversarial loss for spatialization (A vs.\ A+B in Fig.~\ref{fig:DPLM-bar}).
Due to the information bottleneck in the quantized audio codes, not all phase information is sufficiently maintained and reconstructable using a metric loss only. With the addition of adversarial loss, the model is able to generate a plausible phase, resulting in a significant improvement in spatialization quality. In addition, we found that the adversarial loss to be effective at capturing effects such as background noise and reverberation. This can be observed from the spectrograms shown in Figure~\ref{fig:spec}. Because the decoder-only and baseline methods are trained without adversarial loss, the generated speech lacks background noise and reverb details, making the output binaural sound uncanny.

\noindent

\vspace{-5mm}
\section{Conclusions}

We described in detail an the end-to-end binaural speech synthesis system capable of (1) transmitting source monaural speech in the form of compressed speech codes, and (2) synthesizing accurate spatialized binaural speech by conditioning on source and receiver position/orientation information in virtual space. We tested our method on a real-world binaural dataset and found it to be objectively and subjectively superior to a cascade baseline. Finally, we conducted ablation studies to justify various design choices. 

\bibliographystyle{IEEEtran}

\bibliography{references}

\begin{thebibliography}{10}
\providecommand{\url}[1]{#1}
\csname url@samestyle\endcsname
\providecommand{\newblock}{\relax}
\providecommand{\bibinfo}[2]{#2}
\providecommand{\BIBentrySTDinterwordspacing}{\spaceskip=0pt\relax}
\providecommand{\BIBentryALTinterwordstretchfactor}{4}
\providecommand{\BIBentryALTinterwordspacing}{\spaceskip=\fontdimen2\font plus
\BIBentryALTinterwordstretchfactor\fontdimen3\font minus
  \fontdimen4\font\relax}
\providecommand{\BIBforeignlanguage}[2]{{%
\expandafter\ifx\csname l@#1\endcsname\relax
\typeout{** WARNING: IEEEtran.bst: No hyphenation pattern has been}%
\typeout{** loaded for the language `#1'. Using the pattern for}%
\typeout{** the default language instead.}%
\else
\language=\csname l@#1\endcsname
\fi
#2}}
\providecommand{\BIBdecl}{\relax}
\BIBdecl

\bibitem{10.1162/pres.1996.5.3.290}
C.~Hendrix and W.~Barfield, ``{The Sense of Presence within Auditory Virtual
  Environments},'' \emph{Presence: Teleoper. Virtual Environ.}, vol.~5, no.~3,
  p. 290–301, 1996.

\bibitem{binaural-decoder}
A.~Richard, D.~Markovic, I.~D. Gebru, S.~Krenn, G.~A. Butler, F.~Torre, and
  Y.~Sheikh, ``{Neural Synthesis of Binaural Speech From Mono Audio},'' in
  \emph{Proc. ICLR}, 2021.

\bibitem{implicit-HRTF}
I.~D. Gebru, D.~Marković, A.~Richard, S.~Krenn, G.~A. Butler, F.~De~la Torre,
  and Y.~Sheikh, ``{Implicit HRTF Modeling Using Temporal Convolutional
  Networks},'' in \emph{Proc. ICASSP}, 2021, pp. 3385--3389.

\bibitem{richard2022deepimpulse}
A.~Richard, P.~Dodds, and V.~K. Ithapu, ``Deep impulse responses: Estimating
  and parameterizing filters with deep networks,'' in \emph{IEEE International
  Conference on Acoustics, Speech and Signal Processing}, 2022.

\bibitem{soundstream}
N.~Zeghidour, A.~Luebs, A.~Omran, J.~Skoglund, and M.~Tagliasacchi,
  ``{SoundStream: An End-to-End Neural Audio Codec},'' \emph{IEEE/ACM TASLP},
  vol.~30, pp. 495--507, 2022.

\bibitem{LPC}
D.~O'Shaughnessy, ``{Linear predictive coding},'' \emph{IEEE Potentials},
  vol.~7, no.~1, pp. 29--32, 1988.

\bibitem{opus}
J.~Valin, M.~Corporation, K.~Vos, and T.~Terriberry, ``{Definition of the Opus
  Audio Codec},'' 2012.

\bibitem{EVS}
M.~Dietz, M.~Multrus, V.~Eksler, V.~Malenovsky, E.~Norvell, H.~Pobloth,
  L.~Miao, Z.~Wang, L.~Laaksonen, A.~Vasilache, Y.~Kamamoto, K.~Kikuiri,
  S.~Ragot, J.~Faure, H.~Ehara, V.~Rajendran, V.~Atti, H.~Sung, E.~Oh, H.~Yuan,
  and C.~Zhu, ``{Overview of the EVS codec architecture},'' in \emph{Proc.
  ICASSP}, 2015, pp. 5698--5702.

\bibitem{wavenet}
A.~v.~d. Oord, S.~Dieleman, H.~Zen, K.~Simonyan, O.~Vinyals, A.~Graves,
  N.~Kalchbrenner, A.~Senior, and K.~Kavukcuoglu, ``Wavenet: A generative model
  for raw audio,'' \emph{arXiv preprint arXiv:1609.03499}, 2016.

\bibitem{melgan}
K.~Kumar, R.~Kumar, T.~de~Boissiere, L.~Gestin, W.~Teoh, J.~Sotelo,
  A.~de~Br\'{e}bisson, Y.~Bengio, and A.~C. Courville, ``{MelGAN: Generative
  Adversarial Networks for Conditional Waveform Synthesis},'' in \emph{Proc.
  NeurIPS}, vol.~32, 2019.

\bibitem{hifigan}
J.~Kong, J.~Kim, and J.~Bae, ``{HiFi-GAN: Generative Adversarial Networks for
  Efficient and High Fidelity Speech Synthesis},'' in \emph{Proc. NeurIPS},
  vol.~33, 2020, pp. 17\,022--17\,033.

\bibitem{wavenet-speech-coding}
W.~B. Kleijn, F.~S.~C. Lim, A.~Luebs, J.~Skoglund, F.~Stimberg, Q.~Wang, and
  T.~C. Walters, ``{Wavenet Based Low Rate Speech Coding},'' in \emph{Proc.
  ICASSP}, 2018, pp. 676--680.

\bibitem{speech-coding-vqvae-wavenet}
C.~Gârbacea, A.~v. den Oord, Y.~Li, F.~S.~C. Lim, A.~Luebs, O.~Vinyals, and
  T.~C. Walters, ``{Low Bit-rate Speech Coding with VQ-VAE and a WaveNet
  Decoder},'' in \emph{Proc. ICASSP}, 2019, pp. 735--739.

\bibitem{zhen2020cq}
K.~Zhen, M.~S. Lee, J.~Sung, S.~Beack, and M.~Kim, ``{Efficient And Scalable
  Neural Residual Waveform Coding with Collaborative Quantization},'' in
  \emph{Proc. ICASSP}, 2020.

\bibitem{Lyra}
W.~B. Kleijn, A.~Storus, M.~Chinen, T.~Denton, F.~S.~C. Lim, A.~Luebs,
  J.~Skoglund, and H.~Yeh, ``{Generative Speech Coding with Predictive Variance
  Regularization},'' in \emph{Proc. ICASSP}, 2021, pp. 6478--6482.

\bibitem{polyak21_interspeech}
A.~Polyak, Y.~Adi, J.~Copet, E.~Kharitonov, K.~Lakhotia, W.-N. Hsu, A.~Mohamed,
  and E.~Dupoux, ``{Speech Resynthesis from Discrete Disentangled
  Self-Supervised Representations},'' in \emph{Proc. Interspeech}, 2021.

\bibitem{savioja1999creating}
L.~Savioja, J.~Huopaniemi, T.~Lokki, and R.~Väänänen, ``{Creating
  Interactive Virtual Acoustic Environments},'' \emph{{Journal of the Audio
  Engineering Society}}, vol.~47, no.~9, pp. 675--705, 1999.

\bibitem{rendering-localized}
D.~Zotkin, R.~Duraiswami, and L.~Davis, ``{Rendering localized spatial audio in
  a virtual auditory space},'' \emph{IEEE Transactions on Multimedia}, vol.~6,
  no.~4, pp. 553--564, 2004.

\bibitem{natural-sound-rendering}
K.~Sunder, J.~He, E.~L. Tan, and W.-S. Gan, ``{Natural Sound Rendering for
  Headphones: Integration of signal processing techniques},'' \emph{{IEEE
  Signal Processing Magazine}}, vol.~32, no.~2, pp. 100--113, 2015.

\bibitem{surround-by-sound}
W.~Zhang, P.~Samarasinghe, H.~Chen, and T.~Abhayapala, ``{Surround by Sound: A
  Review of Spatial Audio Recording and Reproduction},'' \emph{Applied
  Sciences}, vol.~7, p. 532, 05 2017.

\bibitem{ssl-spatial-audio}
P.~Morgado, N.~Nvasconcelos, T.~Langlois, and O.~Wang, ``{Self-Supervised
  Generation of Spatial Audio for 360\textdegree\ Video},'' in \emph{Proc.
  NeurIPS}, vol.~31, 2018.

\bibitem{2-5d}
R.~Gao and K.~Grauman, ``{2.5 d Visual Sound},'' in \emph{Proc. CVPR}, 2019,
  pp. 324--333.

\bibitem{ssl-spatial-classifier}
Y.-D. Lu, H.-Y. Lee, H.-Y. Tseng, and M.-H. Yang, ``{Self-Supervised Audio
  Spatialization with Correspondence Classifier},'' in \emph{Proc. ICIP}, 2019,
  pp. 3347--3351.

\bibitem{Yang_2020_CVPR}
K.~Yang, B.~Russell, and J.~Salamon, ``{Telling Left From Right: Learning
  Spatial Correspondence of Sight and Sound},'' in \emph{Proc. CVPR}, 2020.

\bibitem{zhou2020sep}
H.~Zhou, X.~Xu, D.~Lin, X.~Wang, and Z.~Liu, ``Sep-stereo: Visually guided
  stereophonic audio generation by associating source separation,'' in
  \emph{Proc. ECCV}, 2020.

\bibitem{vqvae}
A.~van~den Oord, O.~Vinyals, and k.~Kavukcuoglu, ``{Neural Discrete
  Representation Learning},'' in \emph{Proc. NIPS}, 2017, pp. 6306--6315.

\bibitem{vqvae2}
A.~Razavi, A.~Van~den Oord, and O.~Vinyals, ``{Generating Diverse High-fidelity
  Images with VQ-VAE-2},'' in \emph{Proc. NeurIPS}, 2019.

\bibitem{film}
E.~Perez, F.~Strub, H.~De~Vries, V.~Dumoulin, and A.~Courville, ``{FiLM: Visual
  Reasoning with a General Conditioning Layer},'' in \emph{Proc. AAAI},
  vol.~32, no.~1, 2018.

\bibitem{fourier-mapping}
M.~Tancik, P.~Srinivasan, B.~Mildenhall, S.~Fridovich-Keil, N.~Raghavan,
  U.~Singhal, R.~Ramamoorthi, J.~Barron, and R.~Ng, ``{Fourier features let
  networks learn high frequency functions in low dimensional domains},'' in
  \emph{Proc. NeurIPS}, 2020, pp. 7537--7547.

\bibitem{projection-gan}
T.~Miyato and M.~Koyama, ``{c{GAN}s with Projection Discriminator},'' in
  \emph{Proc. ICLR}, 2018.

\bibitem{darwin1999auditory}
C.~Darwin and R.~Hukin, ``Auditory objects of attention: the role of interaural
  time differences.'' \emph{Journal of Experimental Psychology: Human
  perception and performance}, vol.~25, no.~3, p. 617, 1999.

\bibitem{vctk}
C.~Veaux, J.~Yamagishi, and K.~MacDonald, ``{CSTR VCTK Corpus: English
  Multi-speaker Corpus for CSTR Voice Cloning Toolkit},'' 2017.

\bibitem{DPLM}
P.~Manocha, A.~Kumar, B.~Xu, A.~Menon, I.~D. Gebru, V.~K. Ithapu, and
  P.~Calamia, ``{DPLM: A Deep Perceptual Spatial-Audio Localization Metric},''
  in \emph{Proc. Workshop on Applications of Signal Processing to Audio and
  Acoustics (WASPAA)}, 2021, pp. 6--10.

\end{thebibliography}

\end{document}